\documentclass[aps,prl,twocolumn,groupedaddress,nofootinbib,preprintnumbers]{revtex4}

\usepackage{graphicx}
\usepackage{dcolumn}
\usepackage{bm}
\usepackage{latexsym}
\usepackage{amsfonts}
\usepackage{amssymb}
\usepackage{amsmath}
\usepackage{verbatim}

\begin{document}

\preprint{IPMU13-0088}

\title{ SO(3) massive gravity }

\author{Chunshan Lin}
\affiliation{Kavli Institute for the Physics and Mathematics of the Universe (WPI), Todai Institutes for Advanced Study, University of Tokyo, 5-1-5 Kashiwanoha, Kashiwa, Chiba 277-8583, Japan}

\date{\today}

\begin{abstract}
In this paper, we propose a massive gravity theory with 5 degrees of freedom. The mass term is constructed by 3  {\it St\"uckelberg }scalar fields, which respects SO(3) symmetry in the fields' configuration. By the analysis on the linear cosmological perturbations, we found that such 5 d.o.f are free from ghost instability, gradiant instability, and tachyonic instability.
\end{abstract}

\maketitle

{\bf Introduction.~}
The search for a consistent theory of  finite range gravity is a longstanding and well motivated problem. Whether there exists such a consistent extension of general relativity (GR) by a mass term is a basic question of classical field theory. After Fierz and Pauli's pioneering work in 1939 \cite{Fierz1939}, this question has been attracting a great deal of interest. However, its consistency has been a challenging problem for several decades. 

In Fierz and Pauli's model, the GR is extended by a linear mass term. However, such simplest  massive gravity model gives rise to a discontinuity in the observables \cite{vdvz1}\cite{vdvz2}. This problem can be alleviated by nonlinear terms \cite{Vainshtein}. However, since the lack of Hamiltonian constraint and momentum constraint, it ends up with six d.o.f in the gravity sector.  The poincare symmetry in the 3+1 space time implies that a massive spin-2 particle should only contain 5 helicities modes. The rest sixth mode is the so-called Boulware-Deser (BD) ghost \cite{bdghost}, spoiling the stability of the theory. 

Only recently, a non-linear massive gravity theory (which is dubbed as dRGT gravity)  has just been found\cite{deRham:2010ik}\cite{deRham:2010kj},  where the BD ghost is removed by construction in the decoupling limit. It was shown that Hamiltonian constraint and the associated secondary constraint are restored in this theory. As a result, away from decoupling limit, this theory is also free from BD ghost \cite{Hassan:2011hr}. 

With BD ghost free gravity in hand, it is quite necessary to check its cosmological behavior. Interestingly, a self-accelerating solution has been found in ref \cite{Gumrukcuoglu:2011ew}. However, the following up cosmological perturbations analysis revealed a new ghost instability among the rest five d.o.f \cite{Gumrukcuoglu:2011zh}\cite{DeFelice:2012mx}\cite{DeFelice:2013awa}. To avoid the ghost instablity problem, One way to achieve
this would be to relax the FLRW symmetry by deformation of the background \cite{Gumrukcuoglu:2012aa}. Another possibility would be to maintain the FLRW
symmetry and add extra dynamical degrees of freedom to the theory \cite{Huang:2012pe}\cite{D'Amico:2012zv}\cite{Gumrukcuoglu:2013nza}. However, in either case, the theory loses simplicity, and has minor practical application in the cosmology or asrophysics (see \cite{DeFelice:2013bxa} for a review on the cosmology study of dRGT gravity). On the other hand, the acausality problem has been found in ref.  \cite{Deser:2012qx}\cite{Deser:2013eua}, which might strike a deadly blow to dRGT gravity.

By the lesson of dRGT gravity, we learn that if we start from the framwork of breaking the 4 space-time  diffeomorphism invariance, it is very hard to get a healthy 5 degrees of freedom massive spin-2 theory. Inspired by this point, in this paper, we propose a massive gravity thoery by only breaking the 3 spatial diffeomorphism invariance, and keep the time reparameterization invariant,
\begin{eqnarray}\label{rootstone}
t\to t+\xi^0(t,x)~.
\end{eqnarray} 
This theory can also be considered as one of the subcategories of Lorentz violation massive gravity, which was briefly discussed in \cite{Dubovsky:2004sg}. For some other examples of Lorentz violation massive gravity, which break such time reparameterization invariance, see \cite{Blas:2009my}\cite{Comelli:2012vz}\cite{Battye:2013er}\cite{Comelli:2013paa}\cite{Comelli:2013txa}.
 
By taking eq.(\ref{rootstone}) as our starting point, our model only contains 5 degrees of freedom in the gravity sector, which is intrinsically free from BD ghost issue. By adopting the  {\it St\"uckelberg } trick, we introduce 3 scalars, which respect residual SO(3) symmetry in the fields' configuration, to recover the general covariance. 

This paper is organized as follows: Firstly we write down a general action based on the time reparameterization invariance and residual SO(3) symmetry in the scalar fields' configuration.  Then we apply our theory to cosmology. The linear cosmological perturbation analysis reveils 5 healthy degrees of freedom on the perturbation sepctrum.

{\bf Setup }  Taking the time reparameterization invariance and residual SO(3) symmetry  as our building principle, a general action with a masss term can be written as 
\begin{eqnarray}\label{action}
I_g&=&\int d^4x\sqrt{-g}\left\{\frac{M_p^2}{2}R+m_1^2G^{\mu\nu}f_{\mu\nu}\right.\nonumber\\
&&\left.-M_p^2m_2^2\left(c_0+c_1f+c_2f^2+d_2f^{\mu}_{\nu}f^{\nu}_{\mu}+...\right)\right\},
\end{eqnarray}
where $f_{\mu\nu}\equiv \partial_{\mu}\phi^a\partial_{\nu}\phi^b\delta_{ab}~,~a,b=1,2,3$, $f^{\mu}_{\nu}\equiv g^{\mu\rho}f_{\rho\nu}$, $f\equiv f^{\mu}_{\mu}$. Please notice that $G^{\mu\nu}$ is the Einstein tensor and $G^{\mu\nu}f_{\mu\nu}$ is the so called Horndeski term \cite{Horndeski}. The equation of motion is still second derivative.  On the other hand, as for the $M_p^2m_2^2(...)$ part, in principle we can add an infinitely polynomial series inside of round bracket. However, for the simplicity of calculations, we truncate the higher order term, just consider a constant term, a linear term and two quadratic terms in this paper\footnote{The minimal version and its application was discussed in \cite{Lin:2013sja}.}. The stability in the presence of matter content, as well as the stability in other spacetime backgrounds, will be discussed in our future work \cite{clin}.

In the unitary gauge, 
\begin{eqnarray}
\phi^a=\alpha x^a~,~~~f_{\mu\nu}=(0,\alpha^2,\alpha^2,\alpha^2)~,
\end{eqnarray}
where $\alpha$ is an un-normalized quantity, which can be absorbed into the redefinetion of coefficients $c_1,~c_2,~d_2$. Without introducing any ambiguity, we set $\alpha=1$ and $f_{\mu\nu}=(0,~1,~1,~1)$. 

Due to the residual symmetry of time reparameterization invariance, our model only contains 5 degree of freedom in the gravity sector. Thus the BD ghost is absent on the spectrum. 
There is an easy way to observe the stability against BD ghost in our theory.  In terms of ADM formalism, where 
\begin{eqnarray}
ds^2=-N^2dt^2+h_{ij}\left(dx^i+N^idt\right)\left(dx^j+N^jdt\right)~,
\end{eqnarray}
in the unitary gauge, our mass term is a functional of $g^{ij}$ and $f_{ij}$, where $g^{ij}=h^{ij}-\frac{N^i}{N}\frac{N^j}{N}$ and $f_{ij}=\delta_{ij}$. We have four ``would be" Lagrangian multiplier, i.e. $N$ and $N^i$, but there are only 3 combinations of them show up at the mass term. It implies the absence of momentum constraint. However, we can always find a forth combination, i.e. the so called Hamiltonian constraint and its associated secondary constraint,  to eliminate one degree of freedom, and finally we have 5 massive modes in the gravity sector \footnote{See \cite{Gabadadze:2004iv} for an interesting example of Lorentz violation massive gravity, which contains only 2 propagating modes in the gravity sector due to the non-vanishing momentum constraint at linear level.}. 

\begin{figure}
\begin{center}
\includegraphics[width=0.4\textwidth]{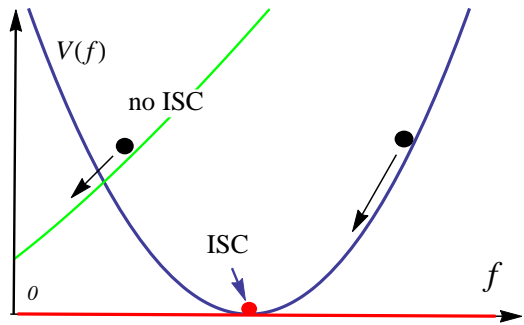}
\end{center}
\caption{Infinitely strong coupling (ISC) issue in the SO(3) massive gravity. The horizontal axes denotes $f\equiv g^{\mu\nu}f_{\mu\nu}$, the blue curve denotes the mass term with  $c_1(3c_2+d_2)<0$, the green curve denotes the mass term with   $c_1(3c_2+d_2)>0$, and the red line, which overlaps with the horizontal axes, denotes an Einstein static universe which exhibit the infinitely strong coupling. Please notice $f$ is positive by definition.}
\label{ISC}
\end{figure}

Before applying our theory to cosmology, we need to clarify the issue concerning strong coupling in Einstein static background. Assume we start from a FRW background, with physical metric,
\begin{eqnarray}\label{FRW}
ds^2=-N^2dt^2+a(t)^2dx^2~,
\end{eqnarray}
and $f=3a^{-2}$ in this case. It is important to notice that for the parameter region $c_1(3c_2+d_2)<0$, when $f$ approaches to the bottom of the potential, one gets an Einstein static universe, provided a proper cosmological constant(see FIG.\ref{ISC}). The helicity 0 mode and helicity 1 mode exhibit the infinitely strong coupling in such Einstein static universe, thus it is not our interest (the detail will be present in \cite{clin}). The condition $c_1(3c_2+d_2)>0$ is required to avoid such problem.

{\bf Cosmological solution}  We choose the FRW ansatz, and the space-time metric can be written as eq.(\ref{FRW}). By taking the variation of the action with respect to the metric $g^{\mu\nu}$, we get such two background Einstein equations,
\begin{eqnarray}\label{freeman1}
3H^2&=&r_2m_2^2\left[c_0+3c_1a^{-2}+3(3c_2+d_2)a^{-4}\right],\\
\frac{\dot{H}}{N}&=&r_2^{2}m_2^2\left[\frac{r_1c_0-3c_1}{3a^{2}}-\frac{(2+r_1a^{-2})(3c_2+d_2)}{a^{4}}\right],\label{freeman2}
\end{eqnarray}
where $r_1\equiv m_1^2/M_p^2$ and $r_2\equiv{M_p^2}/(M_p^2+\frac{m_1^2}{a^2})~.$ In addition to a bare cosmological constant, we can see our mass term contributes a curvature-like term, and a radiation-like term. Noting that since the the SO(3) symmetry in the fields' configuration, the constraint equations of  3 {\it St\"uckelberg } scalars $\phi^a$ are trivially satisfied.

{\bf Scalar perturbations}
We perturb the space-time metric and define the scalar perturbations by 
\begin{eqnarray}
g_{00}&=&-N^2(t)[1+2\phi]~,\\
g_{0i}&=&N(t)a(t)\partial_i\beta~,\\
g_{ij}&=&a^2(t)[\delta_{ij}+2\psi\delta_{ij}+(\partial_i\partial_j-\frac{1}{3}\partial^2)E]~,
\end{eqnarray}
Here we choose the unitary gauge, where $\phi^a=x^a$. After integrating out the non-dynamical degree, the quadratic action of scalar perturbation is 
\begin{eqnarray}\label{2scalar}
I_{scalar}=\frac{M_p^2}{2}\int dtd^3kNa^3\left(\mathcal{K}_{s}\frac{\dot{E}^2}{N^2}-\mathcal{M}_{s}E^2\right)~,
\end{eqnarray}
where 
\begin{eqnarray}
&&\mathcal{K}_s=\nonumber\\
&&\frac{k^{4}\left[a^{2}\left(c_{1}m_{2}^{2}+3H^{2}r_{1}\right)+2m_{2}^{2}\left(3c_{2}+d_{2}\right)\right]}{2r_{2}\left[a^{2}\left(3c_{1}m_{2}^{2}+9H^{2}r_{1}+k^{2}\right)+6m_{2}^{2}\left(3c_{2}+d_{2}\right)+k^{2}r_{1}\right]}\nonumber\\
\end{eqnarray}
The full expression of $\mathcal{M}_s$ is quite bulky, and we are not going to show it here. 

To check if the scalar mode is ghosty, let's substitute  eq.(\ref{freeman1}) into the above formula, and then take the super horizon approximation, we get 
\begin{eqnarray}
\mathcal{K}_s\simeq\frac{1}{6r_2}k^4~.
\end{eqnarray}
The scalar mode is ghost free at super horizon scale as long as $r_2\equiv{M_p^2}/(M_p^2+\frac{m_1^2}{a^2})$ is positive.


To see the situation in the small scale, let's take the sub-horizon approximation, we get 
\begin{eqnarray}\label{skinetic}
\mathcal{K}_s&\simeq&\frac{1}{2}r_1r_2k^2m_2^2c_0+\frac{1}{2}c_1r_2m_2^2k^2\left(1+4r_1a^{-2}\right)\nonumber\\
&&~+\frac{1}{2}\left(3c_2+d_2\right)r_2m_2^2k^2\left(2a^{-2}+5r_1a^{-4}\right).
\end{eqnarray}
At late time epoch where $a\to\infty$, $\mathcal{K}_s>0$ requires
\begin{eqnarray}\label{ghostfree2}
(r_1c_0+c_1)m_2^2>0~,
\end{eqnarray}
and for the early stage where $a\ll 1$,   $\mathcal{K}_s>0$ requires 
\begin{eqnarray}\label{ghostfree3}
(3c_2+d_2)m_2^2>0~.
\end{eqnarray}

In order to check if our theory is free from gradiant instability and tachyonic instability, we define a new canonical variable 
\begin{eqnarray}
\mathcal{E}\equiv\kappa E~,
\end{eqnarray}
where $\kappa$ is defined in the following eq.(\ref{kappa1}) and eq.(\ref{kappa2}). The canonical normalized action can be rewritten in terms of this canonical variable as
\begin{eqnarray}
I_g=\frac{1}{2}\int dtd^3kNa^3\left(\frac{\dot{\mathcal{E}}^2}{N^2}-\omega_s^2\mathcal{E}^2\right)~.
\end{eqnarray}
Under the super horizon approximation, at leading order we have,
\begin{eqnarray}\label{kappa1}
\kappa&\simeq&\frac{k^2\tilde{M}_p}{\sqrt{6}}~,\\
\omega_s^2&\simeq&\frac{m_2^2(4c_1+3r_1c_0)}{a^2}~,
\end{eqnarray}
where $\tilde{M_p}^2\equiv M_p^2+\frac{m_1^2}{a^2}$.  There is no tachyonic instability outside of horizon if $m_2^2(4c_1+3r_1c_0)>0$. 



During late time epoch, under the sub horizon approximation, at the leading order we have
\begin{eqnarray}\label{kappa2}
\kappa&\simeq&k m_2M_p \sqrt{\frac{ \left(c_0 r_1+c_1\right)}{2}}~, \\
\omega_s^2&\simeq&\frac{k^2}{a^2}.
\end{eqnarray}
We can see that during the late time epoch, at leading order the sound speed of scalar mode at subhorizon scale is 1. Although we start from an action break the Lorentz invariance, Lorentz violation effect doesn't show up at the leading oder of our calculation. 

{\bf Vector perturbations} By perfoming the similar approach to vector perturbation, we can also check that the vector mode is also healthy under the same ghost free conditon. Firstly, let's define the vector perturbations of the metric as, 
\begin{eqnarray}
\delta g_{0i}&=&N(t)a(t)S_i~,\\
\delta g_{ij}&=&\frac{1}{2}a^2(t)(\partial_iF_j+\partial_jF_i)~,
\end{eqnarray}
where the vector perturbations satisfy the transverse condition,
\begin{eqnarray}
\partial_iS^i=\partial_iF^i=0~.
\end{eqnarray}

After integrating out the non-dynamical degree, we get the quadratic action of the vector mode as follows,
\begin{eqnarray}\label{2vector}
I_{vector}=\frac{M_p^2}{2}\int dtd^3kNa^3\left(\mathcal{K}_{v}\frac{\dot{F}_i\dot{F}^i}{N^2}-\mathcal{M}_{v}F_iF^i\right),
\end{eqnarray}
where 
\begin{widetext}
\begin{eqnarray}
\mathcal{K}_{v}=\frac{k^2 m_2^2 \left[r_1 \left(a^4 c_0+5 \left(3 c_2+d_2\right)\right)+2 a^2 \left(3 c_2+d_2\right)+c_1 \left(a^4+4 a^2 r_1\right)\right]}{2 r_2 \left[a^4 k^2+4 a^2 m_2^2 \left(a^2 c_1+6 c_2+2 d_2\right)+2 r_1 \left(a^2 k^2+2 m_2^2 \left(a^4 c_0+4 a^2 c_1+5 \left(3 c_2+d_2\right)\right)\right)+k^2 r_1^2\right]}~.
\end{eqnarray}
\end{widetext}
We are not going to show the full expression of $\mathcal{M}_{v}$ since it is too bulky. At super horizon scale, we have
\begin{eqnarray}
\mathcal{K}_v\simeq\frac{k^2}{8r_2}~.
\end{eqnarray}
Again, similar to the scalar case, $r_2>0$ ensures that the kinetic term is positive. At subhorizon scale, 
we have 
\begin{eqnarray}\label{vkinetic}
\mathcal{K}_v&\simeq&\frac{1}{2}c_0r_1r_2m_2^2+\frac{1}{2}c_1r_2m_2^2\left(1+\frac{4r_1}{a^2}\right)\nonumber\\
&&~+\frac{1}{2a^2}(3c_2+d_2)r_2m_2^2\left(2+\frac{5r_1}{a^2}\right).
\end{eqnarray}
By requiring that the vector mode is ghost free at two oppsite limit $a\to\infty$ and $a\ll 1$,  we get exactly the same ghost free condition as in eq.(\ref{ghostfree2})(\ref{ghostfree3}).

In order to check the gradiant instability and tachyonic instability, we write down the canonical normalized action for vector perturbations,
\begin{eqnarray}
I_{vector}=\frac{1}{2}\int dtd^3kNa^3\left(\frac{\dot{\mathcal{F}_i}\dot{\mathcal{F}^i}}{N^2}-\omega_v^2\mathcal{F}_i\mathcal{F}^i\right)~,
\end{eqnarray}
where at leading order,
\begin{eqnarray}
\mathcal{F}_i&\simeq&\frac{\sqrt{2}k\tilde{M_p}}{4}F_i~,\nonumber\\
\omega_v^2&\simeq&\frac{m_2^2(4c_1+3c_0 r_1)}{a^2},~~~~for~k\ll aH~,
\end{eqnarray}
and
\begin{eqnarray}
\mathcal{F}_i&\simeq&\frac{\sqrt{2(c_1+c_0r_1)}}{2}M_pm_2F_i~,\nonumber\\
\omega_v^2&\simeq&\frac{k^2}{a^2},~~~~~~~~for~k\gg aH~.
\end{eqnarray}
The result is quite similar to the scalar perturbations.

{\bf Tensor perturbations} Tensor perturbations on the metric can be defined as 
\begin{eqnarray}
\delta g_{ij}=a(t)^2\gamma_{ij}~,
\end{eqnarray}
where the transverse condition and traceless condition are satisfied, 
\begin{eqnarray}
\partial_i\gamma^{ij}=\gamma^i_i=0~.
\end{eqnarray}

 The quadratic action of the tensor perturbations reads 
\begin{eqnarray}
I_{tensor}=\frac{M_p^2}{4}\int dtd^3kNa^3\left(\mathcal{K}_{T}\frac{\dot{\gamma}_{ij}^2}{N^2}-\mathcal{M}_{T}^2\gamma_{ij}^2\right)~,
\end{eqnarray}
where 
\begin{eqnarray}
\mathcal{K}_T&=&\frac{1}{r_2}~,\\
\mathcal{M}_T^2&=&\frac{k^2}{a^2}\cdot\left(1+\frac{3r_1}{a^2}\right)+M_{GW}^2~.
\end{eqnarray}
Different from GR, the dispersion relation of tensor mode is modified by an effective mass term $M_{GW}^2$.
To see the tensor mode propagating speed, we define a canonical variable to canonical normalize the action, 
\begin{eqnarray}
{\tilde{\gamma}}_{ij}\equiv\sqrt{\frac{2}{r_2}}\cdot\gamma_{ij}~.
\end{eqnarray}
The quadratic action for tensor mode can be rewritten in terms of the canonical variables as,
\begin{eqnarray}
I_{tensor}=\frac{M_p^2}{8}\int dt d^3k Na^3\left[\frac{\dot{\tilde{\gamma}}_{ij}^2}{N^2}-\left(\frac{c_s^2k^2}{a^2}+\tilde{M}_{GW}^2\right)\tilde{\gamma}_{ij}^2\right],\nonumber\\
\end{eqnarray}
where at late time epoch, at leading order 
\begin{eqnarray}
c_s^2&\equiv&\frac{M_p^2+\frac{3m_1^2}{a^2}}{M_p^2+\frac{m_1^2}{a^2}}\simeq 1~,\\
\tilde{M}_{GW}^2&\simeq&\frac{m_2^2(4 c_1 +3 c_0  r_1)}{a^2}~.
\end{eqnarray}

As pointed out in \cite{Gumrukcuoglu:2012wt}, the primary modification due to the mass term of tensor mode is a sharp peak  in the gravitational spectrum. 

{\bf Decoupling limit} The physics in our SO(3) massive gravity becomes extremely simple by adopting the effetive field thoery approach in the decoupling limit,
\begin{eqnarray}
m_2\to0,~M_p\to\infty,~~~keeping ~\left(M_pm_2\right)~fixed~.
\end{eqnarray}
In such decoupling limit, the corresponding action for helicity 0 mode reads schematically as follows,
\begin{eqnarray}
I_{\varphi}=M_p^2m_2^2\int \left(k^2\dot{\varphi}^2-k^4\varphi^2-k^6\varphi^3-k^8\varphi^8-...\right),
\end{eqnarray}
where $\varphi$ is the helicity 0 mode of our Goldstone excitation.
Notice that in order to go to canonical normalization for $\varphi$, we define
\begin{eqnarray}
\varphi^c\equiv (M_pm_2k)\varphi~,
\end{eqnarray}
the canonical normalized action reads as
\begin{eqnarray}
I_{\varphi}=\int\dot{\varphi^c}^2-k^2{\varphi^c}^2-\frac{k^3{\varphi^c}^3}{M_pm_2}-\frac{k^4{\varphi^c}^4}{M_p^2m_2^2}-...
\end{eqnarray}
By comparing the quadratic term, cubic term and quartic term, we can see that  at the energy scale higher than $\Lambda_2=\sqrt{M_pm_2}$,  higher order terms become large and helicity 0 mode gets strongly coupled, thus effective field theory approach breaks down here.  

{\bf Conclusion and discussion} In this short letter, we propose a massive gravity theory based on  SO(3) symmetry and time reparameterization invariance. The time reparameterization invariance ensure that our SO(3) massive gravity is free from the BD ghost. The cost to this virtue is to introduce Lorentz violation. Fortunately, cosmological perturbations analysis tells us such Lorentz violation effect does not show up at the leading order of our calculation, and thus this exotic Lorentz violation effect can be totally negligible at sub horizon scale. On the other hand, by carefully checking the linear perturbations of scalar mode, vector mode and tensor mode, we found that our SO(3) massive gravity is free from ghost instability, gradiant instability, and tachyonic instability. 

By setting the cosmological constant $c_0$ to be zero, and taking the late time limit $a\to\infty$, one gets an asymptotical  Minkowskian background. Similar to the analysis of linear perturbations under the subhorizon approximation, one can easily check that such 5 d.o.f are also healthy in this asymptotical Minkowskian background. 



~~~
\begin{acknowledgments}
The author would like to thank S. Mukohyama, A. E. Gumrukcuoglu, and Y. Wang for the useful discussion and early collaboration. Their contribution is helpful and important for the author to finish this work.  The author also would like to thank G. Gabadadze, X. Gao, K. Hinterbichler, M. Li, N. Tanahashi, A. Tolley, S. Zhou for the useful discussion and comments, and thank S. Deser for the heart warming advice.  This work is supported by the World Premier International Research Center Initiative (WPI Initiative), MEXT, Japan. 
\end{acknowledgments}

\end{document}